\documentclass[aip,jcp,reprint]{revtex4-1}

\usepackage{graphicx}
\usepackage{amsmath}
\usepackage{amssymb}

\usepackage[margin=16pt,font=small,labelfont=bf,justification=raggedright]{caption}
\usepackage[margin=24pt,font=small,labelfont=bf,justification=raggedright]{subcaption}

\newcommand{\Angstrom}{\ensuremath{\mathring{\textnormal{A}}}}
\newcommand{\sub}[1]{\ensuremath{_{\textrm{#1}}}} 
\newcommand{\super}[1]{\ensuremath{^{\textrm{#1}}}} 

\begin{document}

\title{A recipe for free-energy functionals of polarizable molecular fluids}

\author{Ravishankar Sundararaman, Kendra Letchworth-Weaver, and T A Arias}
\affiliation{Cornell University Department of Physics, Ithaca, NY 14853, USA}
\date{\today}


\begin{abstract}
Classical density-functional theory is the most direct approach to
equilibrium structures and free energies of inhomogeneous liquids,
but requires the construction of an approximate free-energy functional 
for each liquid of interest. We present a general recipe for constructing
functionals for small-molecular liquids based only on bulk experimental
properties and \emph{ab initio} calculations of a single solvent molecule.
This recipe combines the exact free energy of the non-interacting system
with fundamental measure theory for the repulsive contribution and
a weighted density functional for the short-ranged attractive interactions.
We add to these ingredients a weighted polarization functional for the long-range correlations
in both the rotational and molecular-polarizability contributions to the dielectric response.
We also perform molecular dynamics calculations for the free energy of cavity formation
and the high-field dielectric response, and show that our free-energy functional adequately
describes these properties (which are key for accurate solvation calculations)
for all three solvents in our study: water, chloroform and carbon tetrachloride.
\end{abstract}

\maketitle

\section{Introduction}

The structure of liquids at the atomic scale critically influences the biological processes 
that are responsible for sustaining life and the chemical reactions that drive modern technology.
However, these systems are difficult to model because liquid phenomena result from
the collective behavior of strongly interacting molecules, making theoretical description challenging.
Classical density-functional theory (CDFT) has proved an effective tool for meeting this challenge.
It has demonstrated great promise for elucidating liquid behavior in diverse systems such as
lipids,\cite{Frink-lipids} polymers,\cite{Wu-polymers,Frink-polymers} confined liquids,\cite{mesoporousCDFT,Wu-poresize}
and electrochemical interfaces.\cite{Wu-DoubleLayer, Wu-Henderson-poresize}

Because of its ability to bridge atomic and  macroscopic length scales,\cite{PorousGeometries, Wu-polymers}
CDFT is inherently a multi-scale theory, but with a rigorous basis.  Providing both
three dimensional microscopic detail of the liquid structure and efficient scaling to large system sizes,
\cite{rigidCDFT,3DTramonto} CDFT is a powerful tool for modeling energy storage and conversion devices.
In fact, CDFT calculations have been used to probe pore size effects\cite{Wu-poresize, Wu-Henderson-poresize}
and thereby holds promise to inform design of better mesoporous supercapacitor,
carbon sequestration,\cite{Gianellis-MPcarbon} and battery\cite{LiSelectrode} materials.  

However, classical DFT alone does not account for details in the electronic structure
which are necessary for describing chemical processes in solvated molecules or surfaces.
The simultaneous need for a quantum mechanical description of the electronic structure
and statistical averaging over the phase space of the liquid
makes theoretical descriptions of such systems challenging.
Standard approaches for the electronic structure of solvated systems range
from the accurate but prohibitively expensive molecular dynamics methods,
either fully \emph{ab initio}\cite{CPMD} or hybrid QM/MM,\cite{QMMM}
to highly empirical but efficient polarizable continuum models.\cite{PCM-Review,PCM-Marzari,NonlinearPCM}
A promising middle ground between these two extremes is joint density-functional theory,\cite{JDFT}
an in-principle exact description of the equilibrium properties of solvated electronic systems
that combines electronic density-functional theory\cite{HK-DFT,KS-DFT} for the solute
with classical density-functional theory for the solvent.

In practice, classical density-functional theory requires an approximation
for the free energy of an inhomogeneous fluid directly in terms of its density.
Excellent approximations for model fluids, particularly the hard-sphere fluid,
have been known for a long time,\cite{SimpleLiquidTheory,WhatIsLiquid}
but the development of accurate functionals for real liquids is still
an area of active research.

Free-energy functional approaches for real liquids fall under two broad classes. The first class of approaches,%
\cite{RISM1,RISM2,WaterFreezingDFT,LischnerH2O,LischnerHCl,WuIntegralEqnCDFT,WaterMultipoleDFT,WaterMultipoleDFT2}
typically based on the weighted-density approximation,\cite{WDA} rely on pair correlation functions
of the fluid from molecular dynamics simulations or neutron and X-ray scattering measurements.
Such methods therefore rely on experimental data available for few liquids at few state points,
or the construction and testing of a pair potential model followed by extensive
molecular dynamics calculations to provide the needed correlation functions.
Consequently, they are not easily applicable to a new solvent or even to previously studied solvents under different conditions.

The second class of approaches\cite{SAFT-Water,WaterSAFTVR,WuBonding,BondedTrimer} require constructing
an effective short-ranged Hamiltonian for the liquid, and then theoretically approximating
the free energy, typically using Wertheim's thermodynamic perturbation theory.\cite{WertheimTPT}
These functionals are easier to extend to other solvents and thermodynamic state points, and are
remarkably accurate for the free energy of cavity formation, one of the two key contributions to solvation
of electronic systems, even though they do not perfectly reproduce the pair correlations of the fluid.
However, most of these functionals do not properly account for dielectric response which is the other
major contribution in solvation and absolutely necessary to the success of a joint density-functional theory.

Recently, we constructed a simplified semi-empirical functional for water\cite{rigidCDFT}
based on the first approach, but without requiring pair correlations as an input.
This `scalar-EOS' functional therefore gains the generality of the second approach,
while improving accuracy for cavity formation as well as dielectric response
by including additional empirical input.
This functional augments a hard sphere fluid with a Lennard-Jones weighted density term,
constrained to the readily measured bulk equation of state and surface tension
of the fluid. These short-ranged terms determine the cavity-formation free energies,
while the nonlinear dielectric response follows from a competition between the
ideal gas entropy and long-range Coulomb interactions between charged sites
on the solvent molecules. 

In this work, we extend the scalar-EOS functional to other molecular fluids, using
weakly-polar chloroform (CHCl\sub{3}) and non-polar carbon tetrachloride (CCl\sub{4})
in contrast to strongly-polar water, to demonstrate the generality of the approach.
Section~\ref{sec:ScalarEOSrecipe} describes the short-ranged part of the scalar-EOS
functional and compares the microscopic cavity-formation free energy for all three
liquids against molecular dynamics simulations. 

The dielectric response of water is dominated by molecular rotation, and 
previous functionals ignore contributions due to electronic and vibrational
polarizability of the molecules, which are much more important for the less polar solvents.
In fact, polarizability and rotations contribute almost equally to the
dielectric constant of chloroform, while the dielectric response of
carbon tetrachloride is entirely due to molecular polarizability.
Section~\ref{sec:DielectricResponse} presents a modified long-range perturbation
scheme that incorporates molecular polarizability contributions, and replaces
the scaled mean-field \emph{ansatz}\cite{LischnerHCl} with a weighted dipole-density approximation
applicable to the rotational and polarization correlations of arbitrary fluids.

Finally, in order to fully specify a scalar-EOS functional for a new liquid,
we must determine some microscopic properties of the solvent. 
In addition to the experimentally measured bulk properties, the free-energy functional 
and its interaction with a solute through joint density functional theory
depend on electron density, charge density, geometry, and nonlocal
susceptibility of a single solvent molecule. Appendix~\ref{sec:AbinitioParameters} details 
a procedure for determining these microscopic properties from a single electronic
density-functional calculation for each new solvent, tabulating the resulting parametrization
for the liquids considered here.

\section{`Scalar-EOS' recipe for free-energy functionals} \label{sec:ScalarEOSrecipe}

\subsection{Overview}

The scalar-EOS approximation for the grand free energy of an inhomogeneous molecular fluid
with probability density $p_\omega(\vec{r})$ of finding a molecule with orientation $\omega\in\textrm{SO}(3)$
at position $\vec{r}$ is
\begin{multline}
\Phi[p_\omega] = \Phi\sub{id}[p_\omega] + \Phi\sub{HS}[N_0] + \int d\vec{r} \bar{N} A\sub{att}(w_A\ast \bar{N},T)\\
	+ \Phi_\epsilon[p_\omega]. \label{eqn:ScalarEOSsummary}
\end{multline}
Here,
\begin{multline}
\Phi\sub{id}[p_\omega] =
	T \int \frac{\textrm{d}\vec{r} \textrm{d}\omega}{8\pi^2} p_\omega(\vec{r})
		\left(\ln\frac{p_\omega(\vec{r})}{N_\textrm{ref}} - 1\right) \\
	+ \sum_\alpha \int \textrm{d}\vec{r} N_\alpha(\vec{r}) (V_\alpha(\vec{r}) - \mu_\alpha)
\end{multline}
is the exact free energy of the ideal gas of rigid molecules with orientation density $p_\omega$,
in external site potentials $V_\alpha$ and at site chemical potentials $\mu_\alpha$, 
where $\alpha$ indexes the independent sites in the molecule.
(See Ref.~\citenum{rigidCDFT} for details.)
$\Phi\sub{HS}[N_0]$ is the excess free energy of a hard sphere fluid with density
$N_0(\vec{r}) = \int \frac{d\omega}{8\pi^2} p_\omega(\vec{r})$ and with an as yet
undetermined hard sphere radius $R\sub{HS}$; this term is approximated using
fundamental measure theory,\cite{RosenfeldFMT,TarazonaFMT,FMTreview}
specifically the White-Bear mark II functional.\cite{WhiteBearFMT_MarkII}

The third term of (\ref{eqn:ScalarEOSsummary}) is a weighted density \emph{ansatz}
for the contributions from intermolecular attractions, with the per-particle
excess Helmholtz energy function $A\sub{att}(N,T)$  constrained to the bulk equation of state
and applied to a weighted combination $\bar{N}$ of the individual site densities $N_\alpha$
convolved with a weight function $w_A$. Instead of deriving this weight function from
the direct correlation function of the liquid as in the weighted-density approximation,\cite{WDA}
we set it to an empirical normalized Lennard-Jones form,
\begin{equation}
w_A(r)= \frac{9}{8\sqrt{2}\pi\sigma^3}
	\begin{cases}
	1/4, & r<2^{1/6}\sigma \\
	\left(\frac{\sigma}{r}\right)^{6}-\left(\frac{\sigma}{r}\right)^{12}, & r\ge2^{1/6}\sigma,
	\end{cases} \label{eqn:LJ-MWF}
\end{equation}
with a range $\sigma$ that we, below, determine from the bulk surface tension of the liquid.
The last term of (\ref{eqn:ScalarEOSsummary}), $\Phi_\epsilon$, accounts for long-range 
orientation-dependent Coulomb interactions and vanishes in the uniform fluid limit.

The fundamental assumption made in the above functional form is the separation of the interactions into
short-ranged ($r^{-6}$ and faster) orientation-averaged and long-range orientation-dependent parts. 
The hard sphere and weighted density terms capture the short-ranged part, while $\Phi_\epsilon$
captures the long-ranged part. We next focus on the short-ranged part, which primarily determines the accuracy
for cavity-formation energies, and explore the long-ranged part responsible for dielectric response
in detail in Section~\ref{sec:DielectricResponse}.

\subsection{Details of short-ranged part}

To fully describe the short-ranged part of the free-energy functional (first three terms of
(\ref{eqn:ScalarEOSsummary})), we must determine the free-energy function
$A\sub{att}(N,T)$ and the density $\bar{N}$ that enter the third term of (\ref{eqn:ScalarEOSsummary}),
the hard sphere radius $R\sub{HS}$, and the attraction range $\sigma$ in (\ref{eqn:LJ-MWF}).
The following paragraphs specify each of these quantities in detail.

First, the per-molecule free energy $A\sub{att}(N,T)$ of a uniform fluid of density $N$ at temperature $T$
can be determined from experimental measurements of the pressure $p(N,T)$ of the bulk fluid.
Specifically, the equation of state $p(N,T)$ determines the grand free-energy density of the uniform fluid
and results in the differential equation $A\sub{att}'(N) = (p(N,T) - p\sub{HS}(N,T))/N^2$, where
$p\sub{HS}(N,T)$ is the Carnahan-Starling equation of state of the hard sphere fluid.\cite{CarnahanStarlingEOS}
For water, the Jeffery-Austin equation of state\cite{JeffAustinEOS}
results in the free-energy function
\begin{multline}
A\sub{att}^{H_2O}(N) =
	\frac{\alpha T}{\lambda b(T)} \ln\frac{1}{1-\lambda b(T) N} - (a\sub{VW}+b^\ast T) N \\
	- 2 T f^{\ast\ast}(T) \frac{1+C_1}{1+C_1 e^{-(N-\rho\sub{HB})^2/\sigma^2}}
		\ln \frac{\Omega_0+\Omega\sub{HB}e^{-\epsilon\sub{HB}/T}}{\Omega_0+\Omega\sub{HB}} \\
	- T \frac{V\sub{HS}N(4-3V\sub{HS}N)}{(1-V\sub{HS}N)^2},
\label{eqn:H2O-Aatt}
\end{multline}
with the numerous constants and functions of temperature as defined in Ref.~\citenum{JeffAustinEOS}.
For the less polar fluids, the generic Tao-Mason equation of state \cite{TaoMasonEOS}
results in the free-energy function
\begin{multline}
A\sub{att}\super{TM}(N) =
	\frac{\alpha T}{\lambda b} \ln\frac{1}{1-\lambda b N} \\
	- T (\alpha - B) \left[ N - A_1 \left( e^{\kappa T_c/T} - A_2\right) \frac{\tan^{-1}(\sqrt{1.8}b^2N^2)}{2\sqrt{1.8}b} \right] \\
	- T \frac{V\sub{HS}N(4-3V\sub{HS}N)}{(1-V\sub{HS}N)^2}.
\label{eqn:TM-Aatt}
\end{multline}
Tao et al. relate the temperature-dependent functions $\alpha(T)$, $b(T)$ and $B(T)$,
as well as the constants $\lambda$, $\kappa$, $A_1$ and $A_2$, to the critical point $(T_c, P_c)$
and acentricity factor, $\omega$, generically for several fluids; see Ref.~\citenum{TaoMasonEOS} for details.
The final terms of (\ref{eqn:H2O-Aatt}) and (\ref{eqn:TM-Aatt}) subtract the free energy corresponding to
the Carnahan-Starling hard-sphere equation of state, with the hard sphere volume $V\sub{HS}=4\pi R\sub{HS}^3/3$,
since $\Phi\sub{HS}$ already accounts for that portion of the free energy in (\ref{eqn:ScalarEOSsummary}).

Next, we consider the weighted combination of densities $\bar{N}$ that determines
the distribution of short-ranged intermolecular attractions (third term of
(\ref{eqn:ScalarEOSsummary}) ) amongst the sites on the molecule.
In the scalar-EOS functional for water,\cite{rigidCDFT} we set the density $\bar{N}$
equal to the scalar moment $N_0$ that enters the hard sphere functional (which also happens to be
oxygen density $N_O$ if the origin of the reference molecular geometry is set to the oxygen atom).
That \emph{ansatz} is suitable for water, since the polarizability is approximately isotropic and it is
reasonable to associate the entire short-ranged interaction to the molecule center (or oxygen site);
this is usually the case for the Lennard-Jones term in pair potential models for water,
including SPC/E,\cite{SPCE} TIP3P\cite{TIP3P} and TIP4P/2005.\cite{TIP4P-2005}

However, for other solvents, the attractive interactions may be dominant for sites far from
the molecular center, such as on the chlorine atoms in chloroform and carbon tetrachloride.
In order to account for this effect without unduly complicating the functional or introducing
additional parameters, we set
\begin{equation}
\bar{N}(\vec{r}) = \sum_\alpha N_\alpha(\vec{r}) \chi_\alpha / \chi\sub{tot}.
\end{equation}
Here, $\chi_\alpha$ is the effective dipole polarizability of each site
and $\chi\sub{tot}$ is the total dipole polarizability of the molecule,
which we obtain from electronic density functional calculations of the
solvent molecule as discussed in appendix~\ref{sec:AbinitioParameters}.
In the dilute limit, this \emph{ansatz} correctly reduces to a $1/r^6$ interaction between each pair
of sites with strength proportional to the product of polarizabilities of the two sites.

\begin{table}
\caption[Hard sphere radii, $R\sub{HS}$, and attraction range, $\sigma$, for scalar-EOS functionals]
{Hard sphere radii, $R\sub{HS}$, set to $R\sub{vdW}$ determined from the equation of state,\cite{SPT-Review,RvdwFluids}
and attraction range, $\sigma$, for which the scalar-EOS functionals reproduce the
bulk surface tension at $T=298$~K (experimental values from Ref. \citenum{CRC-Handbook}).
For water, Ref.~\citenum{rigidCDFT} sets $\sigma = 2R\sub{HS}$ and constrains both
parameters to the surface tension; the resultant $R\sub{HS}$ agrees remarkably
with the standard $R\sub{vdW} = 1.385$~\AA.
\label{tab:RHS}}
\begin{center}
\begin{tabular}{ccc}
\hline\hline
Fluid & $R\sub{HS}$ [\Angstrom] & $\sigma$ [\Angstrom] \\
\hline
H\sub{2}O   & 1.36 & 2.72 \\
CHCl\sub{3} & 2.53 & 2.70 \\
CCl\sub{4}  & 2.69 & 2.78 \\
\hline\hline
\end{tabular}
\end{center}
\end{table}

Finally, we specify the hard sphere radius $R\sub{HS}$, which controls the location
of the first peak in the correlation functions, and the Lennard-Jones diameter $\sigma$,
which controls the range of the intermolecular attraction.
The scalar-EOS recipe for water\cite{rigidCDFT} assumes $\sigma = 2R\sub{HS}$,
as Peng and Yu\cite{LJ-MWFDFT} suggest for the Lennard-Jones fluid, and constrains
the hard sphere radius $R\sub{HS}$ to reproduce the bulk liquid-vapor surface tension.
For liquid water, this results in $R\sub{HS} = 1.36$~\AA, in remarkable agreement
with the standard van der Waals radius $R\sub{vdW} = 1.385$~\AA~ defined in terms
of the effective exclusion volume in the equation of state.\cite{SPT-Review,RvdwFluids}
This correspondence also relies on the validity of attributing the entire short-ranged
term to the molecule center, exactly as in a simple Lennard-Jones fluid.
For chloroform and carbon tetrachloride, assuming $\sigma = 2R\sub{HS}$ and following the
procedure for water results in $R\sub{HS} =$ 2.0 and 2.1~\AA~ respectively,
much smaller than the corresponding $R\sub{vdW} =$ 2.53 and 2.69~\AA.
Proceeding with that \emph{ansatz} then results in a free-energy functional which predicts
the first peak in the pair correlations to be too close, and which underestimates
the free energy of forming microscopic cavities. Therefore, as a general recipe,
we now recommend setting the hard sphere radius $R\sub{HS} = R\sub{vdW}$ determined from the equation of state,
and constrain only the attraction range $\sigma$ to the bulk surface tension.
Table~\ref{tab:RHS} summarizes the $R\sub{HS}$ and $\sigma$ so obtained for water,
chloroform and carbon tetrachloride.

\begin{figure}
\includegraphics[width=\columnwidth]{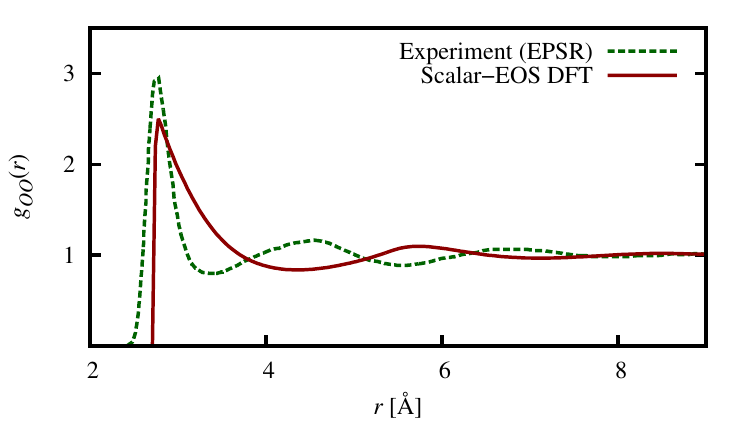}
\caption{Partial $O$-$O$ radial distributions predicted by the scalar-EOS water functional
compared to experimental pair correlations of water from Soper et al.\cite{SoperEPSR}
The position and particle content of the first $g_{OO}$ peak agree reasonably with experiment,
but the remaining structure resembles that of a close-packed hard sphere fluid
rather than a tetrahedrally bonded one.
\label{fig:gOO}}
\end{figure}

\begin{figure}
\centering{\includegraphics[width=\columnwidth]{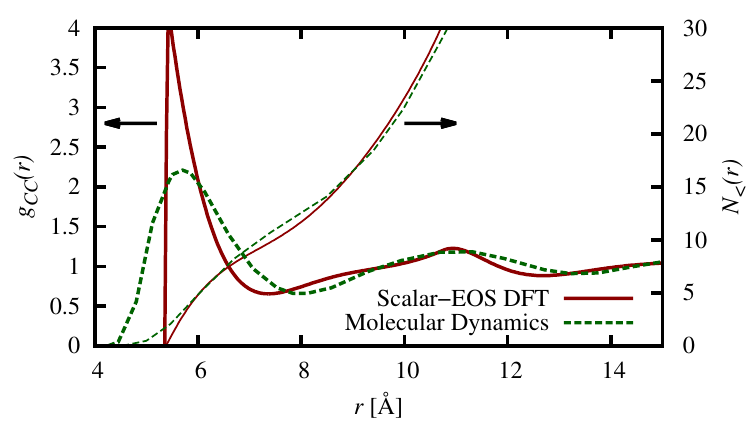}}
\caption{Partial $C$-$C$ radial distributions predicted by the scalar-EOS
carbon tetrachloride functional, compared to molecular-dynamics results.\cite{CCl4-MD}
Despite differences in the shapes of the peaks, their positions and particle contents
(evident from the cumulative distribution $N_<(r)$) agree.
\label{fig:gCC-compare}}
\end{figure}

Figures~\ref{fig:gOO} and \ref{fig:gCC-compare} compare the $O$-$O$ partial radial
distribution in water and the $C$-$C$ distribution in carbon tetrachloride respectively,
predicted by the scalar-EOS functionals against experimental data for water\cite{SoperEPSR}
and molecular dynamics results for CCl\sub{4}.\cite{CCl4-MD}
In both cases, the location of the first peak, which is determined by $R\sub{HS}$,
agrees very well with the reference experimental and molecular dynamics results.
The secondary structure predicted by the functional resembles that of a hard sphere fluid
in both cases, which is in better agreement with the reference data for CCl\sub{4},
than for water, which exhibits tetrahedral structure. In CCl\sub{4}, the density-functional
peaks are narrower and sharper than molecular dynamics because we here have replaced the soft repulsion
by a hard sphere functional. However, the particle contents of the peaks agree quite
well, as can be seen in the cumulative distribution $N_<(r)$ in figure~\ref{fig:gCC-compare}.

\subsection{Cavity formation free energies} \label{sec:ShortRangedResults}

The power of the scalar-EOS approach lies in its capacity to accurately predict
solvation free energies \emph{despite} imperfections in the pair distribution functions.
In the remainder of this section, we focus on the free energy of forming microscopic cavities,
while section~\ref{sec:DielectricResponse} presents the theory and results for the nonlinear dielectric response.
Extensive SPC/E \cite{SPCE} Monte Carlo simulations by Huang et al.
\cite{HardSphereSPCE} provide a reasonable reference estimate for
spherical cavity-formation energies in water, but --- to our knowledge --- similar simulation
results have not yet been published for carbon tetrachloride and chloroform.
Further, the SPC/E model underestimates the surface tension, while the newer TIP4P/2005
model\cite{TIP4P-2005} is more accurate for interfacial energies.
Therefore we estimate the spherical cavity-formation energies using the TIP4P/2005 model
for water. For chloroform we use the model by Lamoureuax et al.\cite{CHCl3-MD} and
for carbon tetrachloride, we use the model by Chang et al.\cite{CCl4-MD}

We perform molecular dynamics calculations for each of the three solvents using
a modified version of LAMMPS \cite{LAMMPS} in which we implemented dipole polarizabilities
for the CHCl\sub{3} and CCl\sub{4} models using classical Drude oscillators.
We use particle-mesh Ewald sums for the Coulomb, as well as $r^{-6}$ interactions.
In each case, we use a periodic cubic simulation box of initial size 32~\AA~ with
the number of molecules set based on the bulk liquid density: 1091 for H\sub{2}O,
245 for CHCl\sub{3} and 203 for CCl\sub{4}. We perform a series of Nose-Hoover NPT calculations at 298~K
and 1~bar using a time step of 1~fs, equilibration time of 200~ps and data collection for 2~ns,
with 33 soft repulsive bias potentials that exclude the liquids from spheres of nominal radii
ranging from 0.3~\AA~ to 9.9~\AA. We then compute the probability $P_>(R)$ of finding a spherical
cavity of radius $R$ or larger in the uniform liquid (from snapshots taken every 0.1~ps)
using umbrella sampling\cite{UmbrellaSampling} and the multiple histogram method,\cite{MultHistMethod}
and thus obtain the free energy of forming a cavity of radius $R$ as $\Delta G(R) = -T\ln P_>(R)$.
See Ref.~\citenum{HardSphereSPCE} for details of the analysis; our calculations differ only
in the model interaction potentials used and in that we used molecular dynamics
instead of Monte Carlo to generate the NPT ensembles, as summarized above.

\begin{figure}
\centering{\includegraphics[width=\columnwidth]{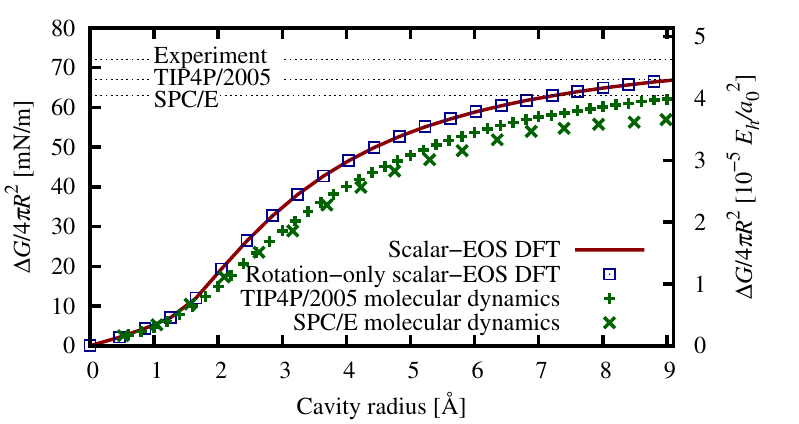}}
\caption[Spherical cavity-formation free energy in H\sub{2}O]
{Free energy per surface area for creating microscopic spherical cavities in water,
as a function of cavity radius,
predicted by the scalar-EOS free-energy functional compared to estimates based on
the SPC/E model (results from Ref.~\citenum{HardSphereSPCE}) and the TIP4P/2005 model.
The dotted lines indicate the bulk surface tension of real water and the predictions of the two models.
\label{fig:sigmavsradius}}
\end{figure}

\begin{figure}
\centering{\includegraphics[width=\columnwidth]{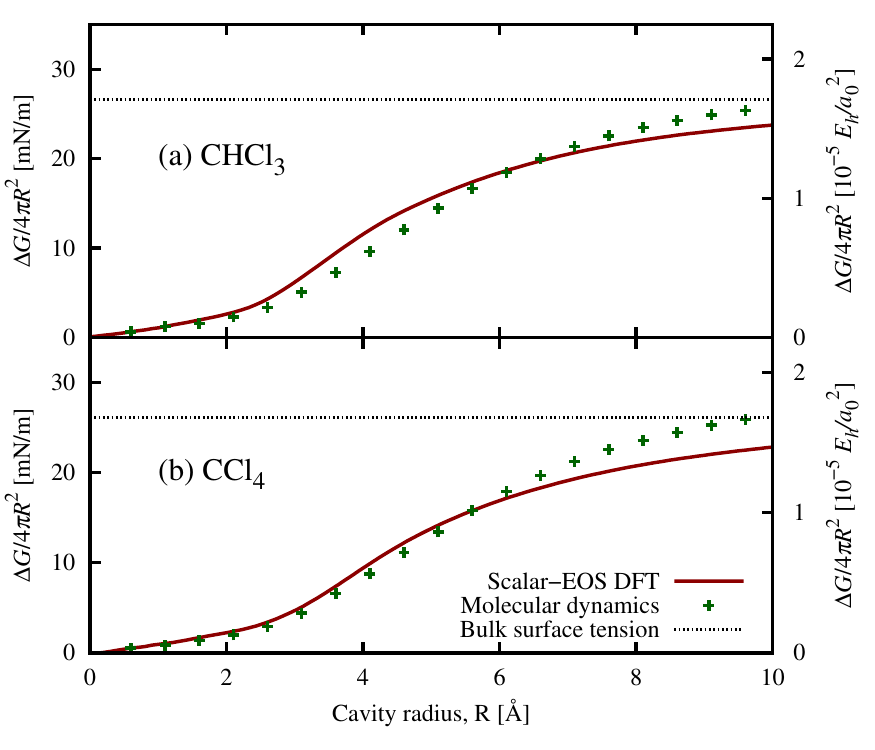}}
\caption[Spherical cavity-formation free energy in CHCl\sub{3} and CCl\sub{4}]
{Free energy per surface area for creating microscopic spherical cavities,
as a function of cavity radius, in
(a) chloroform and (b) carbon tetrachloride, predicted by the scalar-EOS
free-energy functional compared to molecular dynamics estimates.
\label{fig:sigmaVsRadiusCFC}}
\end{figure}

The purpose of the above molecular dynamics calculations is to benchmark the accuracy
of our classical density-functional theory for cavity-formation free energies.
To compute the predictions of our density-functional theories, we implemented the
fluid free-energy functionals in the open-source plane-wave density functional
software, JDFTx.\cite{JDFTx} We calculate the classical density functional estimate of the
cavity-formation free energy by minimizing the free-energy functional (\ref{eqn:ScalarEOSsummary})
in a repulsive external potential that excludes the molecule centers from spherical regions of various radii.
Figure~\ref{fig:sigmavsradius} shows the free energy per surface area required to
form spherical cavities of various radii $R$ in liquid water, as predicted by the
scalar-EOS functional compared against the results of the TIP4P/2005 molecular dynamics
calculations described above as well as the SPC/E results from Ref.~\citenum{HardSphereSPCE}.
We show the density-functional results using two variants of the long-range model from the next section,
one that includes molecular polarizability and the other that only accounts for the rotational response.
The cavity-formation free energy is virtually identical for the two variants of the long-range terms,
justifying our claim that this property is entirely determined by the short-ranged part of the functional.

In Figure~\ref{fig:sigmavsradius}, all estimates agree perfectly for small spheres, where the cavity-formation energy
scales with volume, $\Delta G(R) = N\sub{bulk}T \times 4\pi R^3/3$, rather than with surface area.
Both the SPC/E model and the TIP4P/2005 model underestimate the bulk surface tension, with the
TIP4P/2005 result approximately midway between SPC/E and experiment. The scalar-EOS predictions for the cavity
formation surface energies asymptote to the experimental bulk surface tension for large spheres by construction.
The TIP4P/2005 estimate lies midway between the SPC/E and scalar-EOS values, suggesting that an extrapolation from
TIP4P/2005 and SPC/E to real water based on the bulk values would agree perfectly with the scalar-EOS predictions.
Similarly, Figure~\ref{fig:sigmaVsRadiusCFC} shows that the scalar-EOS predictions for the cavity-formation energies
in chloroform and carbon tetrachloride agree reasonably well with the corresponding molecular dynamics
calculations described above. The scalar-EOS values asymptote to the experimental surface tension,
whereas the molecular dynamics models overestimate the bulk tension by about 2~mN/m for both liquids,
so that our classical DFT predictions are actually more accurate for large radii
than are the predictions of the molecular dynamics models.

\section{Dielectric response including molecular polarizability} \label{sec:DielectricResponse}

The first three terms of the scalar-EOS free-energy functional (\ref{eqn:ScalarEOSsummary})
describe a hard sphere fluid perturbed by short-ranged orientation-averaged attraction.
The final term, $\Phi_\epsilon[p_\omega]$, accounts for long-ranged interactions between charged sites on the solvent molecules,
which in competition with the rotational entropy from the first term, describes the dielectric response.
Ref. \citenum{LischnerH2O} and \citenum{rigidCDFT} both approximate the long-ranged correction
by the mean-field Coulomb interaction with several important modifications.
Here, we examine the motivation for these modifications and propose a new functional form for $\Phi_\epsilon[p_\omega]$
that generalizes to other solvents and includes contributions due to molecular polarizability.

In Ref.~\citenum{rigidCDFT}, the first modification in the mean-field Coulomb interaction attenuates
the Coulomb kernel in reciprocal space at high wave numbers corresponding to the molecular length scale,
in order to minimize spurious intramolecular contributions. We continue to make this approximation, but 
instead motivate it by arguing that the Coulomb kernel may be constructed by minimizing the self-interaction
error for each molecule.  Second, to account for all beyond-mean-field effects, Ref.~\citenum{rigidCDFT} 
introduces an overall scale factor constrained by the bulk linear dielectric constant.
This scale factor cannot account for the differences in correlations in the rotational and polarization responses,
or in the responses of various components in a mixture of fluids. Here, we develop a more natural description
of the beyond-mean-field effects in the form of a weighted polarization-density functional,
which easily generalizes to multiple response contributions or mixtures of fluids.

\subsection{Mean field Coulomb high wave number cutoff}

In the previous functionals,\cite{LischnerH2O,rigidCDFT} the orientation density $p_\omega(\vec{r})$
determines densities $N_\alpha(\vec{r})$ of sites on the solvent molecule with charge $Z_\alpha$,
which then participate in the scaled-mean field Coulomb interaction
\begin{equation}
\Phi_\epsilon[p_\omega] = \frac{A_\epsilon}{2} \sum_{\alpha,\beta} Z_\alpha Z_\beta \int N_\alpha \hat{K} N_\beta
\end{equation}
with an empirical scaling factor $A_\epsilon$ to account for correlations,
and the modified Coulomb kernel $\hat{K}$, specified in reciprocal space as
$\tilde{K}(G) = 4\pi/G^2 \left(1+(G/G_c)^4\right)^{-1}$.
Here, the high wave number cutoff serves to minimize the effects of the Coulomb interaction
at the molecular length scale; this intramolecular contribution primarily results
in a self-interaction error in the mean-field picture.
Lischner et al. set the cutoff wave number $G_c = 0.33\ a_0^{-1}$
by examining the crossover of the direct correlation functions for water,
extracted from neutron diffraction data, from the long-ranged $\sim 1/G^2$
behavior to a more structured short-ranged behavior.\cite{LischnerH2O}
We retain this intuitive picture, but motivate the high wave number attenuation
from an alternate perspective that does not require the direct correlation functions.

The Coulomb kernel at length scales larger than the solvent molecule does not contribute
to the self-interaction error in the mean-field term and, thus, should remain unmodified.
Thus, we set $K(r>2R\sub{vdW}) = 1/r$, since the vdW diameter $2R\sub{vdW}$ is a
reasonable estimate for the typical nearest-neighbor distance in the liquid.
This constraint is implicitly satisfied by the \emph{ansatz} $K(r) = w\sub{MF}(r) \ast 1/r \ast w\sub{MF}(r)$
with a unit-norm short-ranged weight function $w\sub{MF}(r)$ which satisfies $w\sub{MF}(r>R\sub{vdW}) = 0$.
Note that we choose a separable convolution in real-space, or equivalently a separable product
in reciprocal space, to ensure that the resulting interaction may be expressed as the bare Coulomb
interaction acting on sites with spherical charge distributions $Z_\alpha w\sub{MF}(r)$
replacing point charges $Z_\alpha$. This interpretation of evaluating the mean-field term
on an effective charge density, $\rho\sub{MF}(\vec{r}) = \sum_\alpha Z_\alpha w\sub{MF} \ast N_\alpha(\vec{r})$,
then easily generalizes to multiple response channels (such as polarizations) and for mixtures of fluids.

Next, to determine the form for $w\sub{MF}(r)$, we begin by considering a $\delta$-function perturbation
of one of the fluid site densities about the uniform fluid.
The self-interaction error in the above \emph{ansatz} for this configuration is simply the self-energy of the spherical
charge distribution $w\sub{MF}(r)$, up to constants including the magnitudes of the site charge and test perturbation.
Minimizing this self-energy under the constraint $w\sub{MF}(r>R\sub{vdW}) = 0$
results in placing all the charge on the surface of the constraining sphere,
\begin{equation}
w\sub{MF}(r) = \frac{\delta(r - R\sub{vdW})}{4\pi R\sub{vdW}^2},
\label{eqn:mfkernel}
\end{equation}
or equivalently, $\tilde{w}\sub{MF}(G) = j_0(G R\sub{vdW})$ in Fourier space ($j_0$ is a spherical Bessel function).
Intuitively, distributing the charge of each site onto a sphere centered on that site with a radius
that is half the closest intermolecular separation minimizes the intramolecular interaction
while preserving the intermolecular interaction.

\subsection{Inclusion of molecular polarizability effects}

Next, we account for molecular polarizability effects so as to extend the approach to fluids
for which rotations do not dominate the dielectric response to the same extent as in water.
In general, the susceptibility, $\chi(\vec{r},\vec{r}')$, of a molecule to electric potentials due to
electronic polarization and vibrations can be expanded in an eigenbasis $\chi(\vec{r},\vec{r}')
= \sum_i X_i \rho_i(\vec{r}) \rho_i(\vec{r}')$. However, directly employing such a
response in the classical density-functional description would require evaluating the
nonlocal $\chi(\vec{r},\vec{r}')$ operator for each discrete orientation sampled by $p_\omega(\vec{r})$,
making it prohibitively expensive. Molecular dynamics simulations, on the other hand,
demonstrate that it is reasonable to approximate the full nonlocal response by 
independent dipole polarizabilities on each site.\cite{CHCl3-MD,CCl4-MD}

Accordingly, we here employ a nonlocal generalization of this approach and
approximate the response by extended-dipole polarizabilities on each site.
This approximation results in the model susceptibility
\begin{equation}
\chi\sub{model}(\vec{r},\vec{r}') = -\sum_\alpha \chi_\alpha
	\nabla' w_\alpha(|\vec{r}'-\vec{R}_\alpha|) \cdot \nabla w_\alpha(|\vec{r}-\vec{R}_\alpha|) \label{eqn:chiModel}
\end{equation}
for one molecule with sites at positions $\vec{R}_\alpha$, with dipole polarizability strengths
$\chi_\alpha$ and normalized range functions $w_\alpha(r)$. 
In terms of the amplitudes, $\vec{\mathcal{P}}_\alpha$, of the polarization along Cartesian
directions at each site, the potential energy for a polarized state of that molecule is
$\Phi\sub{pol} = \sum_\alpha \mathcal{P}_\alpha^2 / 2\chi_\alpha$
with a corresponding induced charge $\rho(\vec{r}) = \sum_\alpha
\vec{\mathcal{P}}_\alpha \cdot \nabla w_\alpha(|\vec{r}-\vec{R}_\alpha|)$.
In fact, the above susceptibility to electric potential $\phi(\vec{r})$
results from the Euler-Lagrange equation that minimizes the energy
$\Phi\sub{pol} + \int d\vec{r} \phi(\vec{r}) \rho(\vec{r})$.
Appendix~\ref{sec:AbinitioParameters} determines the $\chi_\alpha$ and $w_\alpha(r)$ that best
reproduce the response of a single solvent molecule calculated using electronic density-functional theory.

The nonlocal susceptibility for a molecule assumed above is the most general form that
efficiently generalizes to a fluid specified by site-densities $N_\alpha(\vec{r})$ alone,
rather than depending on the full orientation density $p_\omega(\vec{r})$ in a nontrivial manner.
Within such a framework, the potential energy function generalizes to the functional
\begin{equation}
 \Phi\sub{pol}[\{\vec{\mathcal{P}}_\alpha(\vec{r})\}]
 = \sum_\alpha \int N_\alpha \mathcal{P}_\alpha^2(\vec{r}) / 2\chi_\alpha
\label{eqn:PolPotEnergy}
\end{equation}
in terms of internal variables $\vec{\mathcal{P}}_\alpha(\vec{r})$,
with the corresponding charge density simplifying to
\begin{equation}
\rho\super{pol}(\vec{r}) = -\nabla\cdot \sum_\alpha w_\alpha \ast N_\alpha \vec{\mathcal{P}}_\alpha
\end{equation}
after integrating by parts. This charge density correctly describes the interaction
of the polarization of the fluid with an external electric potential,
but suffers from self-interaction errors when included in the mean-field term.
Following the discussion of the previous section, we can minimize these errors
by distributing the induced charge on a molecule-sized spherical shell.
Therefore, the polarizability contribution to the charge density that enters
the mean field Coulomb interaction is
\begin{equation}
\rho\super{pol}\sub{MF}(\vec{r}) = -\nabla\cdot \sum_\alpha w\sub{MF} \ast N_\alpha \vec{\mathcal{P}}_\alpha,
\end{equation}
with $w\sub{MF}$ given by (\ref{eqn:mfkernel}).

\subsection{Polarization correlations}

Finally, we address beyond-mean-field effects that critically affect the dielectric response.
For simplicity, we first consider a uniform fluid of molecules with density $N$ and
dipole polarizability $\chi$ on each molecule. Assuming only mean-field interactions,
the bulk linear dielectric response of this fluid is described by the free-energy density function
\begin{equation}
\phi(\vec{P}) = \frac{P^2}{2N\chi} + \frac{1}{2}4\pi P^2 - \vec{P}\cdot\vec{D}
\label{eqn:freeenergydensity}
\end{equation}
in terms of the polarization density $\vec{P}$ as the independent variable.
The first term is the potential energy of the molecules in the polarized state,
the second term is the mean-field interaction of the bound charge in the fluid,
which in this case reduces to a long-ranged interaction between sheet charges at the
surface of the dielectric (assuming a parallel-plate capacitor geometry),
and the final term is the interaction with the externally applied field, $\vec{D}$.
Therefore, the equilibrium polarization, $\vec{P} = N\chi\vec{D}/(1+4\pi N\chi)$
results in a net electric field $\vec{E} = \vec{D} - 4\pi\vec{P} = \vec{D}/(1+4\pi N\chi)$,
and hence predicts a dielectric constant of
\begin{equation}
 \epsilon_b = 1 + 4\pi N\chi.
\label{eqn:bulkeps}
\end{equation}
In the dilute or low polarizability limit, this expression is correct to $\mathcal{O}(N\chi)$,
but it is impractically inaccurate for any real fluid.

The Clausius-Mossoti relation accounts for local enhancements in the electric field
interacting with each molecule of the fluid relative to the mean electric field in the medium.
In particular, it places each molecule in a dielectric cavity within which the field
is a factor $C^{CM} = 1 + (\epsilon_b-1)/3$ larger than the mean field.
Solving for $\epsilon_b = \vec{D}/\vec{E}$ from $\vec{P} = N\chi C\vec{E}$ (response to enhanced field)
and $\vec{E} = \vec{D} - 4\pi\vec{P}$ results in the familiar relation,
$\epsilon_b = (1+2\times 4\pi N\chi/3)/(1- 4\pi N\chi/3)$.

Now note that adding a correlation term
\begin{equation}
 (C^{-1}-1) P^2/N\chi
\label{eqn:corrterm}
\end{equation}
to the free-energy function (\ref{eqn:freeenergydensity}) above, implements the response to an enhanced field
by effectively scaling the susceptibility by a factor $C$ and modifying the predicted dielectric constant to
\begin{equation}
 \epsilon_b = 1 + 4\pi N\chi C.
\label{eqn:bulkepsC}
\end{equation}
Further note that theories of the bulk dielectric constant beyond Clausius-Mossoti, such as
the Onsager reaction-field method \cite{DielectricOnsager} or the Kirkwood bond-restriction approach,\cite{DielectricKirkwood}
can all be recast into the above form, but with a different specification of the enhancement factor, $C$.

Within the density-functional perspective, we constrain the enhancement factor
\begin{equation}
 C = (\epsilon_b-1)/(4\pi N\chi)
\label{eqn:Enhancementfactor}
\end{equation}
using (\ref{eqn:bulkepsC}) to reproduce the experimental bulk dielectric constant,
and then generalize the correlation term (\ref{eqn:corrterm}) to the inhomogeneous fluid.
Notice that in the bulk limit, the correlation term (\ref{eqn:corrterm}) can be combined
with the mean-field term (second term of (\ref{eqn:freeenergydensity}))
to obtain a scaled mean-field term, which presents an alternate derivation of the approach
of Refs.~\citenum{rigidCDFT} and \citenum{LischnerH2O}. However, this equivalence no longer holds
in the inhomogeneous fluid, and the correlation functional is more intuitive and tractable
when dealing with real fluids with rotational as well as polarization response.

\begin{table}
\caption[Field enhancement factors for rotational and polarization response]
{Field enhancement factors for the rotational and polarization response of water, chloroform and carbon tetrachloride,
as constrained by (\ref{eqn:CrotCpol}) using the experimental dielectric constants from Ref. \citenum{CRC-Handbook},
compared to that of the Clausius-Mossoti dielectric cavity (labeled by superscript CM).
\label{tab:CrotCpol}}
\begin{center}
\begin{tabular}{c|cc|cc}
\hline
\hline
Fluid & $C\sub{rot}$ & $C\sub{rot}\super{CM}$ & $C\sub{pol}$ & $C\sub{pol}\super{CM}$ \\
\hline
H\sub{2}O & 4.07 & 26.5 & 1.20 & 1.26   \\
CHCl\sub{3} & 2.28 & 1.91 & 1.25 & 1.36 \\
CCl\sub{4} & - & - & 1.26 & 1.38        \\
\hline
\hline
 \end{tabular}
\end{center}
\end{table}

Next, consider the general case of a fluid with a permanent molecular dipole moment $p\sub{mol}$
in addition to site polarizabilities of strength $\chi_\alpha$. The net dipole susceptibility
of the molecule includes a rotational contribution $\chi\sub{rot} = p\sub{mol}^2/(3T)$
as well as a polarization contribution $\chi\sub{pol} = \sum\chi_\alpha$, and
the field-enhancement factors for each contribution would be different in principle.
Within the constraints of available experimental data,
we assume separate enhancement factors for rotations, $C\sub{rot}$, and polarizations, $C\sub{pol}$.
At high frequencies, the rotational response freezes out and the polarizations alone
produce the high frequency dielectric constant, $\epsilon_\infty$, while both
contribute to the static dielectric constant, $\epsilon_b$. Applying (\ref{eqn:bulkepsC})
with $\chi=\chi\sub{pol}$ for $\epsilon_\infty$ and with $\chi=\chi\sub{pol}+\chi\sub{rot}$
for $\epsilon_b$ constrain the enhancement factors
\begin{equation}
C\sub{pol} = \frac{\epsilon_\infty-1}{4\pi N\sub{bulk}\sum_\alpha \chi_\alpha}
\quad\textrm{and}\quad
C\sub{rot} = \frac{\epsilon_b-\epsilon_\infty}{4\pi N\sub{bulk}p\sub{mol}^2/3T}. \label{eqn:CrotCpol}
\end{equation}
Table~\ref{tab:CrotCpol} compares the enhancement factors for the rotational and polarization
contributions with those predicted by the Clausius-Mossoti cavity. The values agree
for the electronic polarizability response for all three fluids, and reasonably so even for
the low dielectric-constant rotational response of chloroform, but are completely different for
the rotational response of water. This conforms to the expectation that the Clausius-Mossoti relation
should be valid for low dielectric constant fluids.

These bulk field enhancement factors do not yet specify a unique functional for the inhomogeneous fluid.
The long-ranged parts of experimental correlation functions are also not sufficiently accurate
to distinguish between the predictions of different long-range functionals that
reduce to the above limit in the uniform fluid, and so we propose the simplest
form to avoid over-parametrization. The potential energy functional for polarization (\ref{eqn:PolPotEnergy})
is explicitly quadratic, exactly as in the bulk linear response limit, and we assume an identical
inhomogeneous form for the correlation functional so that $C\sub{pol}$ simply enhances the site susceptibilities.
For the rotational response, whose `potential energy' is the far more complicated nonlinear ideal gas entropy,
we generalize the correlation term (\ref{eqn:corrterm}) to a weighted polarization-density functional,
and employ the mean-field weight function $w\sub{MF}$ (\ref{eqn:mfkernel}) as an \emph{ansatz} for
the range of the correlations; we describe the final functional form for this term below
(last term of (\ref{eqn:PhiEpsilon})).

\subsection{Net long-range functional}

Collecting the polarizability potential energy (\ref{eqn:PolPotEnergy}), the mean-field interactions and the
correlation functional from the previous sections, the final \emph{ansatz} for the dielectric perturbation functional is
\begin{multline}
\Phi_\epsilon = \sum_\alpha \int d\vec{r} \frac{N_\alpha \mathcal{P}_\alpha^2}{2C\sub{pol}\chi_\alpha}
	+ \frac{1}{2}\int  d\vec{r} \int  d\vec{r}' \frac{\rho\sub{MF}(\vec{r})\rho\sub{MF}(\vec{r}')}{|\vec{r}-\vec{r}'|} \\
	+ \frac{C\sub{rot}^{-1}-1}{N\sub{bulk}p\sub{mol}^2/3T} \int d\vec{r} \bar{P}\sub{rot}^2,
	\label{eqn:PhiEpsilon}
\end{multline}
with the total mean-field effective charge density,
\begin{equation}
\rho\sub{MF} = \sum_\alpha Z_\alpha w\sub{MF} \ast N_\alpha(\vec{r}) - \nabla\cdot \sum_\alpha w\sub{MF} \ast N_\alpha \vec{\mathcal{P}}_\alpha,
\end{equation}
and the weighted rotational polarization-density,
\begin{equation}
\vec{\bar{P}}\sub{rot} = w\sub{MF} \ast \int \frac{d\omega}{8\pi^2} p_\omega(\vec{r}) \omega\circ\vec{p}\sub{mol},
\end{equation}
where $\omega\circ\vec{p}\sub{mol}$ is the dipole moment of the molecule at orientation $\omega$.
Interaction of the fluid with an external electric potential, $\phi(\vec{r})$ takes the form
$\int d\vec{r} \phi(\vec{r}) \rho(\vec{r})$ with the real charge density,
\begin{equation}
\rho(\vec{r}) = \sum_\alpha \rho_\alpha(r) \ast N_\alpha(\vec{r}) - \nabla\cdot \sum_\alpha w_\alpha(r) \ast N_\alpha \vec{\mathcal{P}}_\alpha,
\end{equation}
where $\rho_\alpha(r)$ and $w_\alpha(r)$ are spherical charge-density profiles
and polarizability range functions respectively, for each site.
Appendix~\ref{sec:AbinitioParameters} determines these functions from
electronic density-functional theory calculations using the
parametrizations (\ref{eqn:rhoalpha}) and (\ref{eqn:walpha}) 
for $\rho_\alpha(r)$ and $w_\alpha(r)$.

At this stage, the free-energy functional has the orientation density, $p_\omega$, and the
polarization amplitudes, $\vec{\mathcal{P}}_\alpha$, as independent variables.
The Euler-Lagrange equations for minimizing this functional (including an interaction with
an external electric potential) with respect to $\vec{\mathcal{P}}_\alpha$
show that, at the minimum, all those amplitudes can be expressed as
\begin{equation}
\vec{\mathcal{P}}_\alpha(\vec{r}) = C\sub{pol}\chi_\alpha \left[
	w\sub{MF}(r)\ast\vec{\varepsilon}\sub{MF}(\vec{r})
	- w_\alpha(r) \ast \nabla\phi(\vec{r})
\label{eqn:PolarizabilityatMin}
\right]
\end{equation}
in terms of an auxiliary vector field $\vec{\varepsilon}\sub{MF}(\vec{r})$
(which equals the electric field due to $\rho\sub{MF}$ at the final solution).
In practice, we use the above relation to minimize the free-energy functional with respect to
the independent variables $p_\omega$ (expressed using one of the ideal gas representations
of Ref.~\citenum{rigidCDFT}) and the auxiliary field $\vec{\varepsilon}\sub{MF}(\vec{r})$.
We have revised the rigid-molecular fluid framework\cite{rigidCDFT}
in the open-source electronic density-functional software, JDFTx,\cite{JDFTx}
to include polarizability contributions as detailed above.

\subsection{Results} \label{sec:LongRangedResults}

\begin{figure}
\centering{\includegraphics[width=\columnwidth]{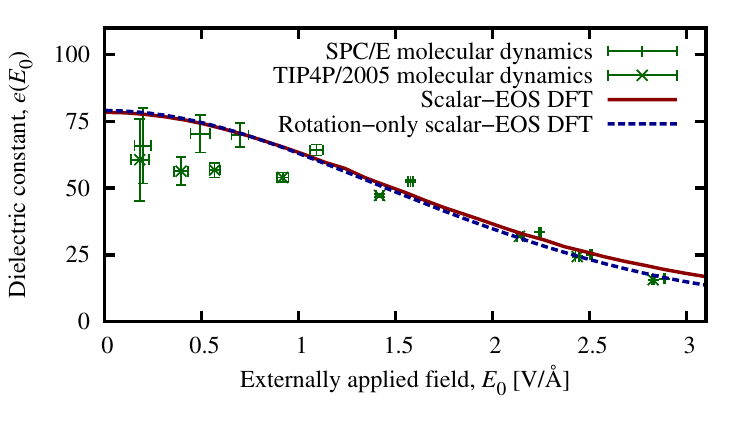}}
\caption[Nonlinear dielectric response versus applied field of H\sub{2}O]
{Nonlinear dielectric response versus applied field of water, predicted by scalar-EOS free-energy functionals
with and without polarizability contributions, compared to molecular dynamics results.
\label{fig:nonlineareps}}
\end{figure}

Figure~\ref{fig:nonlineareps} compares the nonlinear dielectric response of water
predicted by the scalar-EOS free-energy functional with and without molecular polarizability.
The differences are minor and occur only at very high fields for water; adding polarizability slightly
increases the response relative to rotation-only DFT. We also compute the nonlinear response of the SPC/E
and TIP4P/2005 models from 1~ns NPT molecular dynamics calculations using a time step of 1~fs, with various
uniform electric field strengths applied to a periodic cubic simulation box of initial side length 32~\AA.
The other details of the simulations are identical to the ones for the cavity-formation free energy in section~\ref{sec:ShortRangedResults}.
It is interesting to note that the SPC/E model predicts the correct bulk linear dielectric constant and the
TIP4P/2005 model underestimates it, whereas TIP4P/2005 is more accurate for cavity-formation energies.
On the other hand, by construction, the scalar-EOS classical DFT reproduces both of these properties in agreement with experiment.

\begin{figure}
\centering{\includegraphics[width=\columnwidth]{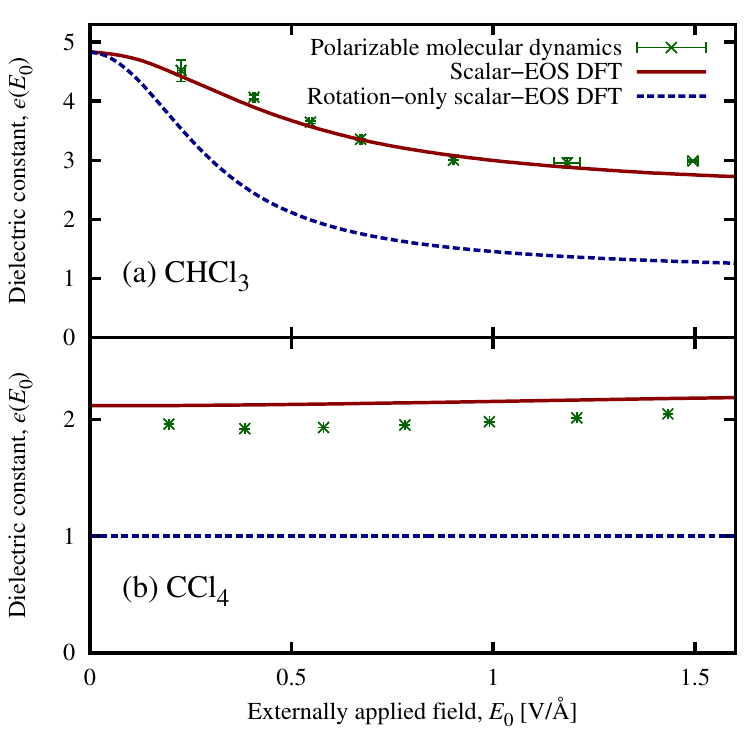}}
\caption[Nonlinear dielectric response versus applied field of CHCl\sub{3} and CCl\sub{4}]
{Nonlinear dielectric response versus applied field of (a) chloroform and (b) carbon tetrachloride,
predicted by scalar-EOS free-energy functionals with and without polarizability contributions,
compared to molecular dynamics results.
\label{fig:nonlinearepsCFC}}
\end{figure}

Figure~\ref{fig:nonlinearepsCFC} presents the analogous results for chloroform and carbon tetrachloride
using the same pair potential models for the molecular-dynamics calculations as in section~\ref{sec:ShortRangedResults}.
Again, the classical DFT predictions agree remarkably well with the molecular dynamics data for both liquids.
However, for these fluids, ignoring polarizability drastically worsens the predicted response.
In chloroform, the predicted dielectric constants using only the rotational response differ significantly
at any finite field and agree in the zero-field limit only by construction.
Carbon tetrachloride has no permanent dipole moment and presents no rotational contribution whatsoever
to the bulk dielectric response. The electronic polarizability contribution increases with field
due to electrostriction (the density of the fluid increases in response to the applied electric field),
and the scalar-EOS DFT with polarizability captures this trend in agreement with molecular dynamics.

\section{Conclusions}

This work presents a general recipe for constructing free-energy functionals
for small molecular liquids by building upon the success of the scalar-EOS
functional for liquid water. The prescribed functional consists of
the exact free energy for the non-interacting system of rigid molecules,
fundamental measure theory for the short-ranged repulsive intermolecular interactions,
a simplified weighted density functional for the short-ranged attractive intermolecular interactions,
and mean-field Coulomb interactions along with a weighted polarization-density correlation functional
for the long-ranged dielectric response. The resulting functional is completely determined
by bulk experimental properties, specifically the equation of state and surface tension,
and by microscopic properties of the solvent molecule that can be derived from electronic
density-functional calculations as detailed in appendix~\ref{sec:AbinitioParameters}.

We test this prescription for three vastly different solvents that range from
highly polar to non-polar: water, chloroform and carbon tetrachloride. We examine
the two key properties that contribute to solvation of electronic systems within 
joint density functional theory: the free energy for forming microscopic cavities,
and the nonlinearities in the dielectric response at high electric fields.
We present reference molecular dynamics simulations of these properties
and demonstrate that our free-energy functionals accurately reproduce them
for all three solvents. In particular, the microscopic cavity-formation free energy
transitions from the volume regime to the surface area regime at the correct length scale.
The rotational dielectric response saturates at the correct electric field scale
and the inclusion of molecular polarizability effects reproduces the correct high-field
behavior including subtle effects such as electrostriction.

In conjunction with an approximation for the interactions between a quantum-mechanical system
(solute or surface) and the fluid\cite{DODFT-Coupling} and a suitable self-consistent
joint minimization scheme, the current prescription for free-energy functionals
will enable joint density-functional theory studies of the solvation of systems described
at the electronic-structure level in equilibrium with small-molecule liquids.
While this work is applicable to a large class of solvents, it would be desirable
to extend such a general prescription to liquids of larger flexible molecules,
mixtures of liquids, electrolytes and ionic liquids in future work.

This work was supported as a part of the Energy Materials Center at Cornell (EMC$^2$),
an Energy Frontier Research Center funded by the U.S. Department of Energy,
Office of Science, Office of Basic Energy Sciences under Award Number DE-SC0001086.
Additional computational time at the Texas Advanced Computing Center (TACC) at the University of Texas
at Austin, was provided via the Extreme Science and Engineering Discovery Environment (XSEDE),
which is supported by National Science Foundation grant number OCI-1053575.

\appendix
\section{Determination of microscopic parameters from \emph{ab initio} calculations} \label{sec:AbinitioParameters}

The bulk equation of state and the surface tension of the fluid completely constrain the short-ranged
part of the scalar-EOS functional. However, the electric response, which dominates the interaction of 
the fluid with electronic systems in solvated electronic-structure calculations, is sensitive to
details of the atomic and electronic structure of the individual molecules.
The geometry, electron and net charge distribution, and susceptibility of each 
constituent solvent molecule affect both the short-range and long-range electric response of the fluid
described in Section~\ref{sec:DielectricResponse}. Although the mean-field Coulomb term internal to the fluid
employs the spherical shell distribution $w\sub{MF}(r)$ to minimize the self-interaction error,
the interaction of the fluid with an external potential employs the charge distribution and
model nonlocal susceptibility of the actual solvent molecule. These quantities are also essential for
determining the interaction of a classical fluid and a quantum-mechanical system within the framework of
joint density-functional theory (JDFT).  Here, we establish the procedure for determining all
these microscopic parameters from electronic density-functional calculations of a single solvent molecule.

A solvent molecule in the liquid environment differs significantly from an isolated or gas phase molecule.
Pair potential models created for molecular dynamics simulations of liquids are calibrated to reproduce 
the thermodynamic properties of the liquid state, but they may capture these differences indirectly.
For example, the dipole moment of the SPC/E model water molecule \cite{SPCE} is 2.35~Debye, in agreement 
with estimates of 2.3-2.5~Debye \cite{pMolFromNonlinearEps} based on cubic susceptibility measurements, 
and in contrast to the gas phase moment of 1.85~Debye. We account for the effect of the surrounding liquid 
by performing the electronic structure calculation of one quantum-mechanical solvent molecule in contact 
with a bath of implicit solvent molecules. To determine the microscopic parameters for a given solvent,
we employ the nonlinear polarizable continuum model,\cite{NonlinearPCM} which approximates solvent effects
in an electronic density-functional calculation of a molecule by surrounding it with a continuum nonlinear dielectric.
In principle, we could obtain the solvent parameters self-consistently within a solvation model which
 includes full microscopic detail.
In practice, however, we find that the parameters determined from a properly constrained and
sufficiently detailed polarizable continuum model are adequate for joint density-functional calculations.

First, we obtain the geometry of the solvent molecule directly from the relaxed nuclear positions,
$\{\vec{R}_\alpha\}$, within a solvated electronic density-functional calculation. 
All \emph{ab initio} calculations were performed using JDFTx\cite{JDFTx}
with the nonlinear polarizable continuum model GLSSA13.\cite{NonlinearPCM}  We employed the
 generalized-gradient approximation \cite{PBE} using a plane-wave basis within
 periodic boundary conditions and a single $k$-point ($\Gamma$) to sample the
 Brillouin zone. Each molecule was computed within a supercell representation
 with a distance of 40~$a_0$ between each periodic image in each direction. 
 All calculations presented employ optimized\cite{Opium} norm-conserving 
Kleinman-Bylander pseudopotentials.\cite{KB} A partial core correction\cite{pcc} was 
required for the Cl pseudopotential. A high plane-wave cutoff energy
 of 70~$E_h$ was chosen so all details in the electron density would be fully 
resolved on a Fourier grid of $(300)^3$ points.
Table~\ref{tab:SolventParameters} shows that the bond lengths and angles, thus obtained,
agree reasonably with popular molecular dynamics models for water, chloroform and carbon tetrachloride.

\begin{table}
\caption[Microscopic solvent parameters from electronic DFT]
{Microscopic solvent parameters for water, chloroform and carbon tetrachloride
from electronic density-functional theory, compared to molecular dynamics models
(Ref.~\citenum{TIP3P} for H\sub{2}O, Ref.~\citenum{CHCl3-MD} for
CHCl\sub{3} and Ref.~\citenum{CCl4-MD} for CCl\sub{4})
wherever applicable.
\label{tab:SolventParameters}}
\begin{center}\begin{tabular}{cccc}
\hline\hline
Solvent & Property & Density-functional & Molecular Dynamics \\
\hline
H\sub{2}O
& $r_{OH}$ & 0.967~\Angstrom & 0.9572~\Angstrom \\
& $\theta_{HOH}$ & 104.2$^\circ$ & 104.52$^\circ$ \\
& $q_O$ & -0.826 & -0.8476 \\
& $q_H$ & +0.413 & +0.4238 \\
& $Z_O\super{el},a_O\super{el}$ & 6.826, 0.37~$a_0$ & - \\
& $d_O\super{el},\nu_O\super{el}$ & 0.52~$a_0$, 0.37~$a_0$ & - \\
& $Z_H\super{el},a_H\super{el}$ & 0.587, 0.35~$a_0$ & - \\
& $d_H\super{el},\nu_H\super{el}$ & 0.0, $2a_H\super{el}/\sqrt{\pi}$ & - \\
& $\chi_O,a_O\super{pol}$ & 3.73~$a_0^3$, 0.32~$a_0$ & - \\
& $\chi_H,a_H\super{pol}$ & 3.30~$a_0^3$, 0.39~$a_0$ & - \\
\hline
CHCl\sub{3}
& $r_{CCl}$ & 1.804~\Angstrom & 1.79~\Angstrom \\
& $r_{CH}$ & 1.091~\Angstrom & 1.1~\Angstrom \\
& $\theta_{HCCl}$ & 107.8$^\circ$ & 107.2$^\circ$ \\
& $q_C$ & -0.256 & -0.175 \\
& $q_H$ & +0.244 & +0.211 \\
& $q_{Cl}$ & +0.004 & -0.012 \\
& $Z_C\super{el},a_C\super{el}$ & 4.256, 0.49~$a_0$ & - \\
& $d_C\super{el},\nu_C\super{el}$ & 0.67~$a_0$, 0.48~$a_0$ & - \\
& $Z_H\super{el},a_H\super{el}$ & 0.756, 0.36~$a_0$ & - \\
& $d_H\super{el},\nu_H\super{el}$ & 0.0, $2a_H\super{el}/\sqrt{\pi}$ & - \\
& $Z_{Cl}\super{el},a_{Cl}\super{el}$ & 6.996, 0.45~$a_0$ & - \\
& $d_{Cl}\super{el},\nu_{Cl}\super{el}$ & 1.01~$a_0$, 0.51~$a_0$ & - \\
& $\chi_C,a_C\super{pol}$ & 6.05~$a_0^3$, 0.36~$a_0$ & 8.84~$a_0^3$, -\\
& $\chi_H,a_H\super{pol}$ & 9.13~$a_0^3$, 0.41~$a_0$ & 0, -\\
& $\chi_{Cl},a_{Cl}\super{pol}$ & 15.8~$a_0^3$, 0.46~$a_0$ & 13.8~$a_0^3$, -\\
\hline
CCl\sub{4}
& $r_{CCl}$ & 1.801~\Angstrom & 1.77~\Angstrom \\
& $q_C$ & -0.980 & -0.1616 \\
& $q_{Cl}$ & +0.245 & +0.0404 \\
& $Z_C\super{el},a_C\super{el}$ & 4.980, 0.61~$a_0$ & - \\
& $d_C\super{el},\nu_C\super{el}$ & 0.53~$a_0$, 0.37~$a_0$ & - \\
& $Z_{Cl}\super{el},a_{Cl}\super{el}$ & 6.755, 0.44~$a_0$ & - \\
& $d_{Cl}\super{el},\nu_{Cl}\super{el}$ & 1.04~$a_0$, 0.52~$a_0$ & - \\
& $\chi_C,a_C\super{pol}$ & 5.24~$a_0^3$, 0.35~$a_0$ & 5.93~$a_0^3$, - \\
& $\chi_{Cl},a_{Cl}\super{pol}$ & 18.1~$a_0^3$, 0.47~$a_0$ & 12.89~$a_0^3$, - \\
\hline\hline
\end{tabular}\end{center}
\end{table}

Next, the electron and nuclear charge densities from the solvated electronic density-functional calculation
are expanded as a sum of spherical contributions around each atom of the solvent molecule, $\sum_\alpha\rho_\alpha(r)$.
The exponential tails of the solvent electron density overlap with the corresponding tails
of the solute and therefore affect the solute-solvent interaction terms\cite{DODFT-Coupling}
in joint density-functional theory.\cite{JDFT} To a certain extent, we can choose functions to represent the 
electron and nuclear charge densities within the core region to optimize representability
on a Fourier grid without changing the interaction energies. The contribution to the charge density
from the core electrons and the nuclei has norm $Z_\alpha\super{nuc}$ (determined by the pseudopotential
choice for valence/core separation). This charge is confined to the interior regions of the molecule
(does not overlap with the cores of other molecules). Thus, to ensure optimum Fourier resolvability,
we smooth it with a Gaussian distribution of standard deviation
$\sigma_\alpha\super{nuc} = R_{0\alpha}/6$. These distributions then become zero to numerical precision
at the atomic vdW radius $R_{0\alpha}$, which is a reasonable estimate for the typical approach distance
of that site to any other atom. 

\begin{figure}
  \centering
        \begin{subfigure}{\columnwidth}
                \includegraphics[width=\columnwidth]{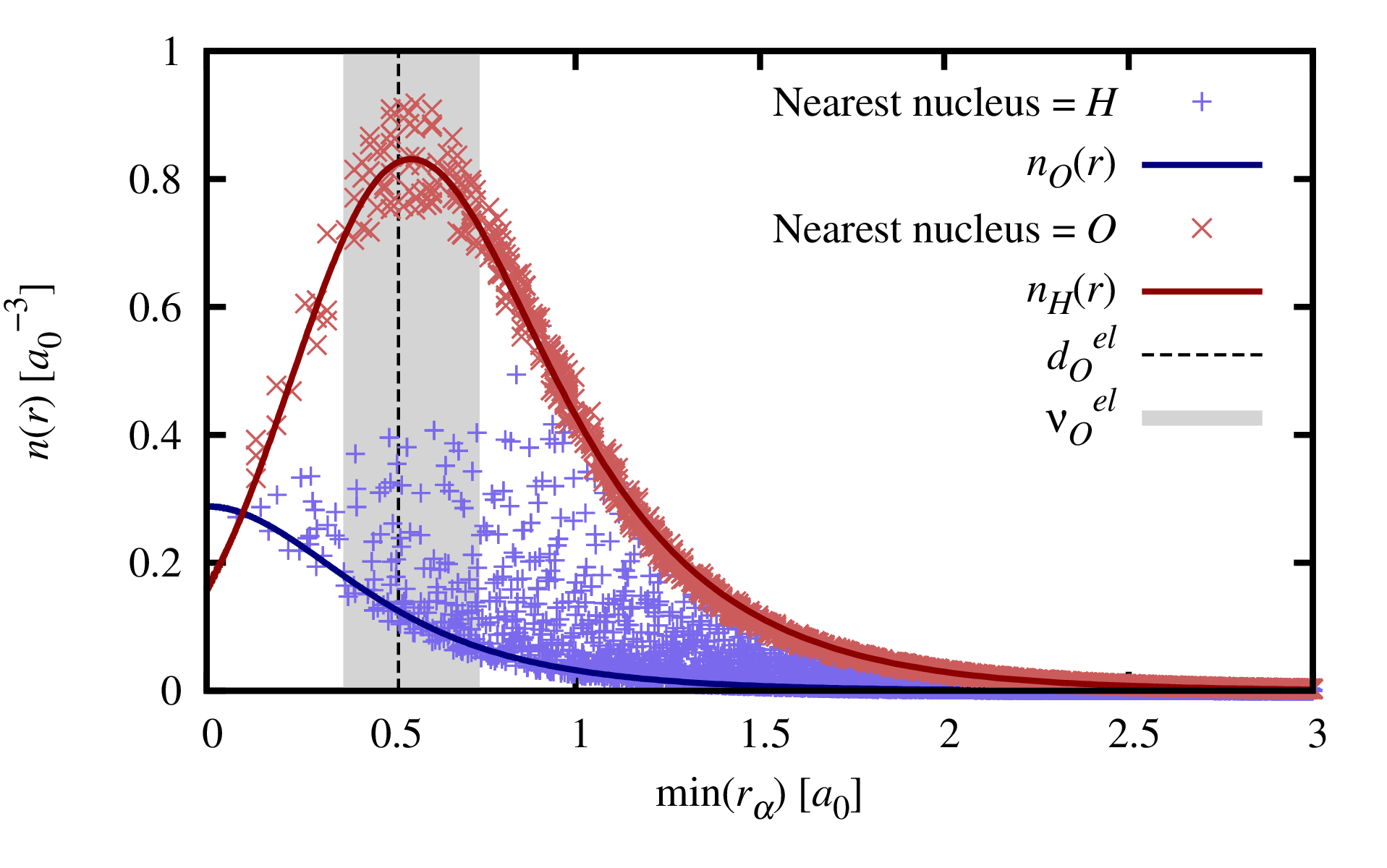}
                \caption{ Individual site density models $n_{O,H}(\vec{r})$ with parameters
                  given in Table~\ref{tab:SolventParameters}
                  compared to the valence electron density $n(\vec{r})$ at each point in space.
                  Parameters $d_{O}\super{el}$ and  $\nu_{O}\super{el}$
                  are indicated by the dotted line and the width of the gray box.}
                \label{fig:OHelecdens}
        \end{subfigure}

        \begin{subfigure}{\columnwidth}
                \includegraphics[width=\columnwidth]{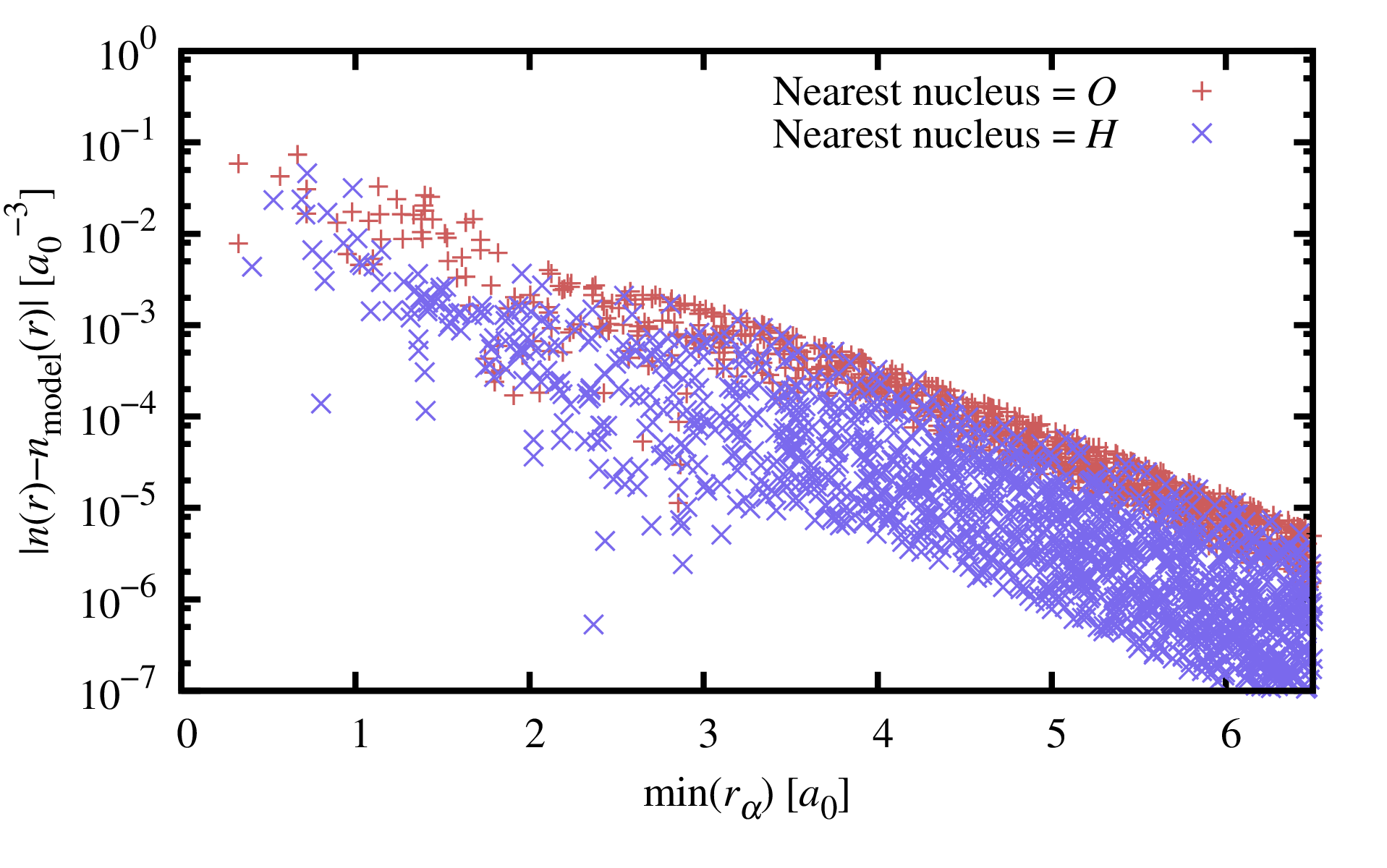}
                \caption{ Error in the spherical density decomposition at each point in space}
                \label{fig:OHresiduals}
        \end{subfigure}

\caption{Spherical decomposition of the water molecule's electron density versus the distance to the nearest nucleus.}
\label{fig:nFitWater}
\end{figure}

However, those functions which monotonically decrease from a maximum at the site center
(such as a simple exponential or Gaussian) and only provide
one degree of freedom (such as a decay width $a$) are not sufficiently accurate to
describe the valence electron densities of the solvent molecules.
When a width is chosen to reproduce only the asymptotic density tails, these monotonic functional
forms disagree significantly with the valence electron densities, even at atomic radii
 beyond the van der waals radius. These issues are compounded for atoms represented
 within the pseudopotential framework, where the core electrons are missing (as in Figure~\ref{fig:nFitWater}(a)). 
For each site, we thus require a function which smoothly increases
 away from the origin to account for the missing core electrons, yet has the correct asymptotic
 exponentially decaying behavior $\propto e^{-r/a}$.  For the valence electron density component attributed to site $\alpha$, 
the (unnormalized) function 
\begin{equation}
f_\alpha(\vec{r})=\mbox{erfc}\left(\frac{r_\alpha-d_\alpha\super{el}}
{\nu_\alpha\super{el}}\right)e^{-r_\alpha/a_\alpha\super{el}}
\end{equation}
 meets these criteria, with $r_\alpha = |\vec{r}-\vec{R}_\alpha|$ as the distance
 from nucleus $\alpha$, $a_\alpha\super{el}$ as the exponential decay length scale,
 $d_\alpha\super{el}$ determining the location of the peak, and $\nu_\alpha\super{el}$
 determining the peak width. See Figure \ref{fig:nFitWater}(a) for an illustration of
the physical meanings of these parameters $\{ a_\alpha\super{el}$, $d_\alpha\super{el}$, $\nu_\alpha\super{el} \}$. 
For the hydrogen atom (or any other atom where all core electrons are included explicitly),
we fix $d_\alpha\super{el}=0$ and $\nu_\alpha\super{el}=\frac{2 a_\alpha\super{el}}{\sqrt{\pi}}$
to create a function which is cuspless at the origin.
We then fit the full valence electron density of the solvent molecule $n(r)$ to the model form
\begin{equation}
n\sub{model}(\vec{r}) = \sum_\alpha
	\frac{ Z_\alpha\super{el}f_\alpha(\vec{r})}{\int_Vd\vec{r}f_\alpha(\vec{r})},
\end{equation}
where $Z_\alpha\super{el}$ is the norm associated with the electron density component at site $\alpha$,
and the denominator is present to normalize the function $f_\alpha(\vec{r})$ over the calculation unit cell volume $V$.

We constrain the norms of all sites $Z_\alpha\super{el}$ to match the lowest multipole moments of the solvent molecules,
employing as many moments as necessary to constrain them (up to dipole for water,
quadrupole for chloroform, and octupole for carbon tetrachloride).
We then select the parameters $\{ a_\alpha\super{el}$, $d_\alpha\super{el}$, $\nu_\alpha\super{el} \}$
to minimize the least-squares residual $\int d\vec{r} |n(\vec{r})-n\sub{model}(\vec{r})|^2$.
The core regions are included in the fit, but have a smaller effect because
there are far more values of $\vec{r}$ in the exponential tails. 
Figures~\ref{fig:nFitWater}(a)~and~(b) compare the valence electron density 
and the site-spherical model for water. Note that the electron density is reproduced well
 in both the intermediate and tail regions, and the residual in the core regions has zero 
multipole moments to high order by construction and therefore does
not contribute to the electric interaction with another non-overlapping molecule.
Table~\ref{tab:SolventParameters} shows the electron density fit parameters
and the implied site charges $q_\alpha = Z_\alpha\super{nuc} - Z_\alpha\super{el}$. 

Our site charges agree reasonably with those of common pair potentials for the highly polar liquid water,
whose thermodynamic properties are sensitive to these parameters in molecular dynamics simulations,
and to a lesser extent, for the weakly polar liquid chloroform. However, in non-polar fluids,
the bulk thermodynamic properties do not constrain the multipole moments, since the magnitude of
the Coulomb interaction is insignificant compared to the magnitude of the dispersion interaction.
Thus, unsurprisingly, the empirically determined molecular dynamics site charges
for carbon tetrachloride \cite{CCl4-MD} differ significantly from our \emph{ab initio} values.
In fact, the octupole moment of our CCl\sub{4} model is 13.2~Debye-\Angstrom\super{2} in much better
agreement with the experimental value of $(15\pm 3)$~Debye-\Angstrom\super{2} \cite{CCl4-Octupole},
compared to 0.5~Debye-\Angstrom\super{2} for the model of Ref.~\citenum{CCl4-MD}.

From these fits, the total charge density kernel for interactions
of the classical fluid with external electric potentials is then given by
\begin{multline}
  \rho_\alpha(r) = \frac{Z_\alpha\super{nuc}}{(\sigma_\alpha\super{nuc}\sqrt{2\pi})^3}  \exp\left(\frac{-r_\alpha^2}{2(\sigma_\alpha\super{nuc})^2}\right) \\
	- 	\frac{ Z_\alpha\super{el}f_\alpha(\vec{r})}{\int_Vd\vec{r}f_\alpha(\vec{r})}.
\label{eqn:rhoalpha}
\end{multline}

Finally, the electronic polarizability $\chi(\vec{r},\vec{r}')$ in Kohn-Sham electronic density functional theory
is formally related to the susceptibility of the corresponding non-interacting system,
\begin{equation}
\chi\sub{NI}(\vec{r},\vec{r}') = -4\sum_{c,v} \frac{\psi_c(\vec{r})\psi_v^\ast(\vec{r})\psi_c^\ast(\vec{r}')\psi_v(\vec{r}')}{\epsilon_c-\epsilon_v},
\end{equation}
by $\hat{\chi}^{-1} = \hat{\chi}\sub{NI}^{-1} - \delta^2 E_{HXC}[n] / \delta n^2$.
Here, $(\psi_v,\epsilon_v)$ and $(\psi_c,\epsilon_c)$ are occupied and unoccupied Kohn-Sham orbital-eigenvalue
pairs respectively, and $E_{HXC}[n]$ is the sum of the Hartree term and the exchange-correlation functional.
In practice, we compute a large number of unoccupied Kohn-Sham eigenpairs of the solvated solvent molecule,
compute $\hat{\chi}$ from $\hat{\chi}\sub{NI}$ as a dense matrix in the occupied-unoccupied basis ($\{\psi_c^\ast(\vec{r})\psi_v(\vec{r})\}$),
and then diagonalize $\hat{\chi}$ to obtain an eigen-expansion $\chi(\vec{r},\vec{r}') = \sum_i X_i \rho_i(\vec{r}) \rho_i(\vec{r}')$.
We find that 1000 unoccupied orbitals and 500 eigenvectors in the final expansion results
in better than 1~\% convergence in the total dipole polarizability of the molecule.
The penultimate column of table~\ref{tab:CrotCpol} shows that the calculated isotropic linear dipole
polarizabilities are within 5~\% of the experimental values \cite{CRC-Handbook} for all three liquids.

The classical density functional requires the polarizability in the model form, $\hat{\chi}\sub{model}$ given by (\ref{eqn:chiModel}).
In order to properly represent the exponential tail regions with a smooth core region,
we pick a cuspless exponential form
\begin{equation}
w_\alpha(r) = \frac{r + a_\alpha\super{pol}}{32\pi(a_\alpha\super{pol})^4} \exp\left(\frac{-r}{a_\alpha\super{pol}}\right)
\label{eqn:walpha}
\end{equation}
for the normalized range functions. We then fit the site polarizability strengths, $\chi_\alpha$,
and widths, $a_\alpha\super{pol}$, to minimize the residual $\textrm{Tr}\left((\hat{K}(\hat{\chi}-\hat{\chi}\sub{model}))^2\right)$
which effectively measures the error in the screening operator $\hat{\epsilon}^{-1} = 1 - \hat{K}\hat{\chi}$,
where $\hat{K}$ is the Coulomb operator. Table~\ref{tab:SolventParameters} lists the thus obtained
polarizability parameters for all three solvents. Note that the width parameters for a particular species
are relatively similar in different solvents, while the strengths differ. Also, the empirically-fit 
polarizability parameters used for each atom in the pair-potential models \cite{CHCl3-MD,CCl4-MD}
compare reasonably to our \emph{ab initio} parameters for $C$ and $Cl$, but neglect the response at the $H$ site.

The procedures outlined above make specific choices for residuals and functional forms for fitting
which are, of course, by no means unique. However, the parametrization developed here approximates
the full \emph{ab initio} charge distributions and susceptibilities well for the studied solvents.
As such, the above prescription enables the construction of a free-energy functional for a new
solvent of interest, without requiring extensive experimental or molecular dynamics data.

%

\end{document}